\documentclass[twocolumn,epjc3,final]{svjour3}  
\smartqed  
\RequirePackage{graphicx}
\RequirePackage{mathptmx}      
\RequirePackage{latexsym,amssymb}
\RequirePackage[colorlinks,citecolor=blue,urlcolor=blue,linkcolor=blue]{hyperref}
\journalname{Eur. Phys. J. C}

\def\prd{Physical Review D }

\def\apj{Astroph. J. }
\def\apjl{Astroph. J. }
\def\mnras{Mon. Not. R. Astron. Soc.}

\begin{document}

\title{The blackholic quantum
}


\author{J. A. Rueda\thanksref{e1,addr1,addr2,addr3,addr4,addr5} \and R. Ruffini\thanksref{e2,addr1,addr2,addr6} 
}

\thankstext{e1}{e-mail: jorge.rueda@icra.it}
\thankstext{e2}{e-mail: ruffini@icra.it}

\institute{ICRANet, Piazza della Repubblica 10, I--65122 Pescara, Italy \label{addr1}
           \and
           ICRA, Dipartimento di Fisica, Sapienza Universit\`a di Roma, P.le Aldo Moro 5, I--00185 Rome, Italy \label{addr2}
           \and
           ICRANet-Ferrara, Dipartimento di Fisica e Scienze della Terra, Universit\`a degli Studi di Ferrara, Via Saragat 1, I--44122 Ferrara, Italy \label{addr3}
           \and
           Dipartimento di Fisica e Scienze della Terra, Universit\`a degli Studi di Ferrara, Via Saragat 1, I--44122 Ferrara, Italy \label{addr4}
           \and
           INAF, Istituto de Astrofisica e Planetologia Spaziali, Via Fosso del Cavaliere 100, 00133 Rome, Italy \label{addr5}
           \and
           INAF, Viale del Parco Mellini 84, 00136 Rome  Italy\label{addr6}
}

\date{Received: date / Accepted: date}

\maketitle

\begin{abstract}
We show that the high-energy emission of GRBs originates in the \textit{inner engine}: a Kerr black hole (BH) surrounded by matter and a magnetic field $B_0$. It radiates a sequence of discrete events of particle acceleration, each of energy ${\cal E} = \hbar\,\Omega_{\rm eff}$, the \textit{blackholic quantum}, where $\Omega_{\rm eff} =4(m_{\rm Pl}/m_n)^8(c\,a/G\,M)(B_0^2/\rho_{\rm Pl})\Omega_+$.
Here $M$, $a=J/M$, $\Omega_+=c^2\partial M/\partial J=(c^2/G)\,a/(2 M r_+)$ and $r_+$ are the BH mass, angular momentum per unit mass, angular velocity and horizon; $m_n$ is the neutron mass, $m_{\rm Pl}$, $\lambda_{\rm Pl}=\hbar/(m_{\rm Pl}c)$ and $\rho_{\rm Pl}=m_{\rm Pl}c^2/\lambda_{\rm Pl}^3$, are the Planck mass, length and energy density. {Here and in the following use CGS-Gaussian units}. The time\-scale of each process is $\tau_{\rm el}\sim \Omega_+^{-1}$, {along the rotation axis, while it is much shorter off-axis owing to energy losses such as synchrotron radiation}. We show an analogy with the Zeeman and Stark effects, properly scaled from microphysics to macrophysics, that allows us to define the \textit{BH magneton}, $\mu_{\rm BH}=(m_{\rm Pl}/m_n)^4(c\,a/G\,M)e\,\hbar/(M c)$. We give quantitative estimates for GRB 130427A adopting $M=2.3~M_\odot$, {$c\, a/(G\,M)= 0.47$ and $B_0= 3.5\times 10^{10}$~G}. Each emitted \textit{quantum}, {${\cal E}\sim 10^{37}$~erg}, extracts only {$10^{-16}$} times the BH rotational energy, guaranteeing that the process can be repeated for thousands of years. The \textit{inner engine} can also work in AGN as we here exemplified for the supermassive BH at the center of M87.
\end{abstract}

\section{Introduction}\label{sec:1}

The GeV radiation in long GRBs is observed as a continuous, \textit{macroscopic} emission with a luminosity that, in the source rest-frame, follows a specific power-law behavior: for instance the $0.1$--$100$~GeV rest-frame luminosity of GRB 130427A observed by Fermi-LAT is well fitted by $L=A~t^{-\alpha}$, $A = (2.05\pm 0.23) \times 10^{52}$~erg~s$^{-1}$ and $\alpha=1.2 \pm 0.04$ \cite{2018arXiv180305476R}. We have there shown that the rotational energy of a Kerr BH is indeed sufficient to power the GeV emission. From the global energetics we have determined the BH parameters, namely its mass $M$ and angular momentum per unit mass $a=J/M$ and, from the change of the luminosity with time, we have obtained the slowing-down rate of the Kerr BH. We have applied this procedure to the GeV-emission data of $21$ sources. For GRB 130427A, we obtained that the BH initial parameters are $M\approx 2.3~M_\odot$ and {$c\,a/(G\,M)\approx 0.47$} \cite{2019ApJ...886...82R}.



One of the most extended multi-messenger campaign of observation in the field of science, ranging from ultra high-energy photons GeV/TeV (MAGIC) and MeV radiation (Swift, Fermi, Integral, Konus/WIND and UHXRT satellites) and to fifty optical observatories including the VLT, has given unprecedented details data on GRB 190114C. An in-depth time-resolved spectral analysis of its prompt emission, obtaining the best fit of the spectrum, and repeating it in successive time iterations with increasingly shorter time bins has been presented in \cite{2019arXiv190404162R}. It turns out that the spectra are self-similar and that the gamma-ray luminosity, expressed in the rest-frame, follows a power-law dependence with an index $-1.20\pm 0.26$, similar to the one of the GeV luminosity.

These data have offered us the first observational evidence of the moment of BH formation and, indeed, it clearly appears that the high-energy radiation is emitted in a sequence of elementary events, each of {$10^{37}$~erg}, and with an ever increasing repetition time from {$10^{-14}$ to $10^{-12}$~s} \cite{2019ApJ...886...82R}.

We have shown that this emission can be powered by what we have called the \textit{inner engine} \cite{2019ApJ...886...82R,2019arXiv190404162R,2019arXiv190511339R}: a Kerr BH immersed in a magnetic field $B_0$ and surrounded by matter. This \textit{inner engine} naturally forms in the binary-driven hypernova (BdHN) scenario of GRBs \cite{2012ApJ...758L...7R,2014ApJ...793L..36F,2016ApJ...833..107B,2018ApJ...852..120B}. The BdHN starts with the supernova explosion of a carbon-oxygen star that forms a tight binary system with a neutron star companion. The supernova ejecta produces a hypercritical accretion process onto the neutron star bringing it to the critical mass point for gravitational collapse, hence forming a rotating BH. The Kerr BH, in presence of the magnetic field inherited from the neutron star, induces an electromagnetic field that is described by the Wald solution \cite{1974PhRvD..10.1680W}. The BH is surrounded by matter from the supernova ejecta that supply ionized matter that is accelerated to ultrarelativistic energies at expenses of the BH rotational energy. This model has been applied to specific GRB sources in \cite{2019ApJ...886...82R,2019arXiv190404162R,2019ApJ...883..191R}.

We here show that the GRB high-energy (GeV/TeV) radiation is indeed better understood within this scenario and that in particular: 1) it originates near the BH and 2) it is extracted from the BH rotational energy by \textit{packets}, \textit{quanta} of energy, in a number of finite discrete processes. We show that it is indeed possible to obtain the \textit{quantum of energy} of this elementary process: ${\cal E} = \hbar \Omega_{\rm eff}$, where $\Omega_{\rm eff}$ is proportional to the BH angular velocity, $\Omega_+$, and the proportionality constant depends only on fundamental constants. The timescale of the elementary process is shown to be $\tau_{\rm el}\sim \Omega_+^{-1}$. Quantitatively speaking, initially {${\cal E}\approx 10^{37}$~erg} and $\tau_{\rm el}$ is shorter than microseconds.

This elementary process is not only finite in energy but it uses in each iteration only a small fraction of the BH rotational energy which can be as large as $E_{\rm rot}\sim 10^{53}$~erg. As we shall see, this implies that the repetitive process, in view of the slowing-down of the BH, can lasts thousands of years. The considerations on the \textit{inner engine} apply as well to the case of AGN and we give a specific example for the case of M87*, the supermassive BH at the center of the M87.

\section{The \textit{inner engine} electromagnetic field structure}\label{sec:2}

The axisymmetric Kerr metric for the exterior field of a rotating BH, in Boyer-Lindquist coordinates $(t,r,\theta,\phi)$, can be written as \cite{1968PhRv..174.1559C}:
\begin{eqnarray}
\label{Kmetric}
ds^2 &=& -\left( 1- \frac{2\hat{M} r}{\Sigma} \right) (c dt)^2-\frac{ 4\hat{a}\,\hat{M}\,r\sin^2\theta }{\Sigma} c dt d\phi +\frac{\Sigma}{\Delta} dr^2\nonumber\\ &+&\Sigma d\theta^2+ \left[r^2+\hat{a}^2+\frac{2r\hat{M}\hat{a}^2\sin^2\theta}{\Sigma}\right]\sin^2\theta d\phi^2,
\end{eqnarray}
where $\Sigma=r^2+\hat{a}^2\cos^2\theta$ and $\Delta=r^2-2\hat{M}r+\hat{a}^2$. The (outer) event horizon is located at $r_+=\hat{M}+\sqrt{\hat{M}^2-\hat{a}^2}$, where $\hat{M}=G\,M/c^2$ and $\hat{a}=a/c$, being $M$ and $a=J/M$, respectively, the BH mass and the angular momentum per unit mass. {Quantities with the hat on top are in geometric units.}

Denoting by $\eta_\mu$ and $\psi_\mu$, respectively, the timelike and spacelike Killing vectors, the electromagnetic four-potential of the Wald solution is $A_\mu = \frac{1}{2} B_0\,\psi_\mu + \hat{a}\,B_0 \eta_\mu$, where $B_0$ is the test magnetic field value \cite{1974PhRvD..10.1680W}. The associated electromagnetic field (in the Carter's orthonormal tetrad), for {parallel} magnetic field and BH spin, is:
\begin{eqnarray}
   E_{\hat{r}} &=& \frac{\hat{a} B_0}{\Sigma} \left[r\sin^2\theta-\frac{\hat{M} \left(\cos ^2\theta+1\right) \left(r^2-\hat{a}^2 \cos ^2\theta \right)}{\Sigma}\right],\\
    E_{\hat{\theta}} &=& \frac{\hat{a} B_0}{\Sigma}\sin\theta \cos\theta \sqrt{\Delta},\\ 
  B_{\hat{r}} &=& -\frac{B_0}{\Sigma} \cos\theta \left(-\frac{2 \hat{a}^2 \hat{a} r \left(\cos^2\theta+1\right)}{\Sigma }+\hat{a}^2+r^2\right),\\
  B_{\hat{\theta}} &=& \frac{B_0 r}{\Sigma}\sin\theta \sqrt{\Delta}.
 \end{eqnarray}

\section{Energetics and timescale of the elementary process}\label{sec:3}

The electrostatic energy {gained by an electron (or proton for the antiparallel case) when accelerated} from the horizon to infinity, along the rotation axis, is
\begin{equation}\label{eq:epsilonp}
\epsilon_e = -e A_\mu \eta^\mu |_{\infty} + e A_\mu \eta^\mu |_{\rm r_+}= e\,\hat{a}\,B_0=\frac{1}{c}e\,a\,B_0,
\end{equation}
%
where we have used that $\psi_\mu \eta^\mu = 0$ and $\eta_\mu\eta^\mu \to -1$ along the rotation axis, and $\eta_\mu\eta^\mu = 0$ on the horizon \cite{1974PhRvD..10.1680W}. 

The electric field {for $\theta=0$, and} at the horizon, $E_+$, is \cite{2019ApJ...886...82R}:
\begin{equation}\label{eq:Eh}
    |E_+| = \frac{1}{2}\frac{\hat{a}}{\hat{M}} B_0 = \frac{1}{2}\frac{c J}{G\,M^2} B_0 \approx \frac{1}{c}\Omega_+ r_+ B_0,
\end{equation}
where the last expression is accurate for $\hat{a}/\hat{M}\lesssim 0.5$ \cite{2019ApJ...886...82R}, and it evidences the \textit{inducting} role of the BH angular velocity
\begin{equation}\label{eq:Omegah}
    \Omega_+= \frac{\partial M c^2}{\partial J} = c\frac{1}{2}\frac{\hat{a}/\hat{M}}{r_+}.
\end{equation}

Using Eq.~(\ref{eq:Eh}), Eq.~(\ref{eq:epsilonp}) can be written as
\begin{equation}\label{eq:epsilonp3}
    \epsilon_e \approx e\,|E_+|\,r_+\approx \frac{1}{c} e\,r_+^2\,\Omega_+\,B_0.
\end{equation}

It is worth to note that this angular frequency can be related to the {energy gained} timescale of the elementary process:
\begin{equation}\label{eq:timescale}
    \tau_{\rm el} = \frac{\epsilon_e}{e |E_+| c}\approx \frac{r_+}{c} = \frac{\hat{a}/\hat{M}}{2 \Omega_+},
\end{equation}
that is the {characteristic acceleration time of the particle along the BH rotation axis}. Thus, this is the longest timescale for the elementary process and it happens on the rotation axis where no (or negligible) radiation losses occur. {This is relevant for the emitting power of ultrahigh-energy charged particles leading to ultrahigh-energy cosmic rays}. Off-polar axis, the {charged} particles emit e.g. synchrotron radiation {at GeV energies, in a much shorter timescale of the order of $10^{-14}$~s} (see \cite{2019ApJ...886...82R} for details).

The total electric energy available for the \textit{inner engine} elementary process is \cite{2019ApJ...886...82R}:
\begin{equation}\label{eq:Etotal}
    {\cal E}\approx \frac{1}{2} |E_+|^2 r_+^3 = \frac{1}{4}\frac{\hat{a}}{\hat{M}} \frac{r_+ \Omega_+}{c} r^3_+B_0^2,
\end{equation}
where in the last equality we have used Eqs.~(\ref{eq:Eh}) and (\ref{eq:Omegah}).
%

\section{The quantum of energy for GRBs}\label{sec:4}

We recall that in a BdHN the BH is formed from the collapse of a neutron star when it reaches the critical mass, $M_{\rm crit}$, by accreting the ejected matter in the supernova explosion of a companion carbon-oxygen star \cite{2012ApJ...758L...7R,2014ApJ...793L..36F,2015ApJ...812..100B,2015PhRvL.115w1102F,2016ApJ...833..107B,2018ApJ...852..120B,2019ApJ...871...14B}. Thus, for the GRB case we can adopt $r_+\sim 2\,G\,M/c^2$ and $M = M_{\rm crit}\sim m_{\rm Pl}^3/m_n^2$,
%
%
%
where $M_{\rm crit}$ is accurate within a factor of order unity; $m_{\rm Pl}=\sqrt{\hbar c/G}$ and $m_n$ are the Planck and neutron mass. With this, the energy per proton (\ref{eq:epsilonp3}) can be written in the \textit{quantized} form:
\begin{equation}
    \epsilon_e = \hbar\,\omega_p,\qquad
    \omega_p\equiv \frac{4\,G}{c^4}\left(\frac{m_{\rm Pl}}{m_n}\right)^4\,e\,B_0\,\Omega_+.
   \label{eq:epsilonp4}
\end{equation}
%

Following the above steps for $\epsilon_e$, we can also write Eq.~(\ref{eq:Etotal}) in the \textit{quantized} form:
%
%
\begin{equation}
    {\cal E} = \hbar\,\Omega_{\rm eff},\qquad
    \Omega_{\rm eff}\equiv 4\left(\frac{m_{\rm Pl}}{m_n}\right)^8\left(\frac{\hat{a}}{\hat{M}}\right)\,\left(\frac{B_0^2}{\rho_{\rm Pl}}\right)\Omega_+,
    \label{eq:Equantum}
\end{equation}
%
where $\rho_{\rm Pl}\equiv m_{\rm Pl} c^2/\lambda_{\rm Pl}^3$ and $\lambda_{\rm Pl}=\hbar/(m_{\rm Pl} c)$ are the Planck energy-density and length. The quantities in parenthesis are dimensionless; e.g.~$B_0^2$ is an energy density as it is $\rho_{\rm Pl}$. Each discrete process extracts a specific amount of the BH rotational energy set by the \textit{blackholic quantum} (\ref{eq:Equantum}).

%

\section{Specific quantitative examples}\label{sec:5}

Concerning quantitative estimates, let us compute the main physical quantities of the \textit{inner engine} for the case of GRB 130427A \cite{2019ApJ...886...82R}. We have there estimated that, an \textit{inner engine} composed of a newborn BH of $M\approx 2.3~M_\odot$, {$\hat{a}/\hat{M} = 0.47$ and $B_0=3.5\times 10^{10}$~G}, can explain the observed GeV emission. {We} recall that the \textit{inner engine} parameters in \cite{2019ApJ...886...82R} were determined at the end of the prompt emission (at $37$~s rest-frame time). At that time, the observed GeV luminosity is $L_{\rm GeV}\approx 10^{51}$~erg~s$^{-1}$. {The timescale of synchrotron radiation expected to power this emission was found to be $t_c \sim 10^{-14}$~s (to not be confused with $\tau_{\rm el}$), which implies an energy ${\cal E}\sim L_{\rm GeV}\times t_c = 10^{37}$~erg, consistent with the blackholic quantum estimated here for the above \textit{inner engine} parameters (see Table~\ref{tab:parameters})}.

The elementary, discrete process introduced here can also be at work in AGN where the time variability of the high-energy GeV-TeV radiation appears to be emitted on subhorizon scales (see \cite{2014Sci...346.1080A} for the case of M87*). Thus, we also show in Table~\ref{tab:parameters} the physical quantities for an AGN, which can be obtained from the expressions in section~\ref{sec:3}. We adopt as a proxy M87*, so $M\approx 6\times 10^9~M_\odot$ (e.g. \cite{2013ApJ...770...86W}), and we assume respectively for the BH spin and the external magnetic field, $\hat{a}/\hat{M} = 0.9$ and $B_0=50$~G. The magnetic field has been fixed to explain the observed high-energy luminosity which is ${\rm few}\times 10^{42}$~erg~s$^{-1}$ (e.g.~\cite{2009ApJ...707...55A,2015MNRAS.450.4333D}).
%
%

\begin{table}
    \centering
    \begin{tabular}{c|c|c}
    \hline & GRB (130427-like) & AGN (M87*-like)\\
    \hline
       $\tau_{\rm el}$ & $2.21\times 10^{-5}$~s & $0.49$~day\\
       $\epsilon_e$~(eV) & {$1.68\times 10^{18}$} & $1.19\times 10^{19}$\\
       ${\cal E}$~(erg) & {$4.73\times 10^{36}$} & $5.19\times 10^{47}$\\
       $\dot{{\cal E}}$~(erg/s) & {$2.21\times 10^{41}$} & $1.22\times 10^{43}$\\
       \hline
    \end{tabular}
    \caption{\textit{Inner engine} astrophysical quantities for GRBs and AGN. {The power reported in the last row is the one to accelerate ultrahigh-energy particles, i.e. $\dot{{\cal E}} = {\cal E}/\tau_{\rm el}$.} In both cases the {parameters (mass, spin and magnetic field) have} been fixed to explain the observed high-energy ($\gtrsim$~GeV) luminosity.}
    \label{tab:parameters}
\end{table}

This shows that the energy of the \textit{blackholic quantum} is finite and is a very small fraction of the BH rotational energy: for GRBs we have $E_{\rm rot}\sim 10^{53}$~erg and {${\cal E}/E_{\rm rot}\approx 10^{-16}$} and for AGN ${\cal E}/E_{\rm rot}\approx 10^{-13}$. This guarantees that the emission process has to occur following a sequence of the elementary processes. Under these conditions, the duration of the repetitive sequence, $\Delta t\sim(E_{\rm rot}/{\cal E})\tau_{\rm el}$, can be of thousands of years, in view of the slowing-down of the BH leading to an ever increasing value of $\tau_{\rm el}$ \cite{2019ApJ...886...82R} (while ${\cal E}$ holds nearly constant). 

\section{The \textit{black hole magneton}}\label{sec:6}

It is interesting to show the analogy of the above result with the case of an atom placed in an external electric or magnetic field for which its energy levels suffer a shift, respectively, from the Stark or Zeeman effect (see e.g. \cite{1965qume.book.....L}). 

In the case of the Zeeman effect, the energy shift is:
\begin{equation}\label{eq:zeeman}
    \Delta \epsilon_Z = \mu_B B_0,\qquad \mu_B\equiv e \frac{\hbar}{2 m_e c},
\end{equation}
where $\mu_B$ is the Bohr magneton. 
Indeed, by using $\Omega_+\approx c (\hat{a}/\hat{M})/(4 G M/c^2)$, and introducing $\mu_{\rm BH}$, the \textit{BH magneton},
\begin{equation}
    \mu_{\rm BH}\equiv \left(\frac{m_{\rm Pl}}{m_n}\right)^4\left(\frac{\hat{a}}{\hat{M}}\right)e \frac{\hbar}{M c},
\end{equation}
the particle energy (\ref{eq:epsilonp4}) can be written as
\begin{equation}
    \epsilon_e = \mu_{\rm BH} B_0,
\end{equation}
which adds an unexpected deeper meaning to $\epsilon_e$.

In the Stark effect, the energy shift is given by
\begin{equation}\label{eq:stark}
    \Delta \epsilon_S = e\,|E_+|\,r_B,
\end{equation}
where $r_B = \hbar^2/(m_e e^2)$ is the Bohr radius. This expression can be directly compared with the first equality in Eq.~(\ref{eq:epsilonp3}).

\section{A direct application to the electron}\label{sec:7}

The use of the Wald solution overcomes the conceptual difficulty of explaining the origin of the charge in BH electrodynamics. Indeed, an effective charge of the system can be expressed as \cite{1974PhRvD..10.1680W,2019ApJ...886...82R}
\begin{eqnarray}\label{eq:Qeff}
  Q_{\rm eff}= \frac{G}{c^3}2\,J\,B_0,
\end{eqnarray} 
which is not an independent parameter but, instead, it is a derived quantity from the BH angular momentum and the magnetic field $B_0$. These quantities become the free parameters of the electrodynamical process and therefore the concept of the BH charge is not anymore a primary concept.

The effective charge $(\ref{eq:Qeff})$ can be also expressed in terms of $M$, $J$ and the magnetic moment $\mu$ as:
\begin{equation}\label{eq:Qeff2}
    Q_{\rm eff} = \frac{M c}{J}\mu,
\end{equation}
where we have used the computation of the Geroch-Hansen multipole moments \cite{1970JMP....11.2580G,1974JMP....15...46H} performed in 
\cite{1974PhRvD..10.1680W}. Assuming the electron spin $J_e=\hbar/2$ and $Q_{\rm eff}=e$, the magnetic moment becomes the Bohr magneton. But more interestingly, if we adopt the angular momentum and magnetic moment of the electron, then we obtain that the derived effective charge (\ref{eq:Qeff2}) becomes indeed the electron charge:
\begin{equation}
    Q_{\rm eff} = \frac{m_e\,c}{J_e} \mu_B = \frac{2 m_e\,c}{\hbar}\frac{\hbar}{2 m_e\,c}\,e = e.
\end{equation}

\section{Conclusions}\label{sec:8}

We recall:

1)
That in addition of being exact mathematical solutions of the Einstein equations, BHs are objects relevant for theoretical physics and astrophysics as it was clearly indicated in ``Introducing the BH'' \cite{1971PhT....24a..30R}. 

2)
That the mass-energy of a Kerr-Newman BH, established over a few months period ranging from September 17, 1970, to  March 11, 1971 in \cite{1970PhRvL..25.1596C,1971PhRvD...4.3552C,1971PhRvL..26.1344H}, can be simply expressed by
\begin{eqnarray}
\label{eq:Mbh}
M^2 &=& \frac{c^2 J^2}{4 G^2 M^2_{\rm irr}}+\left(\frac{Q^2}{4 G\, M_{\rm irr}}+M_{\rm irr}\right)^2,\\
S &=& 16 \pi G^2\,M^2_{\rm irr}/c^4,\quad \delta S = 32 \pi  G^2\,M_{\rm irr} \delta M_{\rm irr}/c^4  \geq 0,
\end{eqnarray}
where $Q$, $J$ and $M$ are the three independent parameters of the Kerr-Newman geometry: charge, angular momentum and mass. $M_{\rm irr}$ and $S$ are, respectively, the derived quantities representing the irreducible mass and the horizon surface area. 

3)
That for extracting the Kerr BH rotational energy the existence of the Wald solution \cite{1974PhRvD..10.1680W} was essential  \cite{2019ApJ...886...82R,2019arXiv190404162R,2019ApJ...883..191R}.

From the observational point of view, the time-resolved spectral analysis of GRB 130427A \cite{2018arXiv180305476R,2019ApJ...886...82R} and GRB 190114C \cite{2019arXiv190404162R} clearly points to the existence of self-similarities in the Fermi-GBM spectra, to the power-law in the GeV luminosity of the Fermi-LAT and to a discrete emission of elementary impulsive events of {$10^{37}$~erg}. The timescale {of the emission is,} {on the rotation axis $\sim 10^{-6}$~s, leading to ultrahigh-energy particles contributing to cosmic rays, and off-axis of $\sim 10^{-14}$~s, leading to GeV-TeV radiation \cite{2019ApJ...886...82R}}.

{
Extrapolating these considerations from a BH to an electron, we showed that the electron charge turns out to be a derived quantity, a function of the electron's angular momentum and magnetic moment, with the electron's mass and the speed of light considered as fundamental constants.
}

The definition, the formulation of the equation and the identification of the mechanism of the process of emission of the \textit{blackholic quantum} has become a necessity and it is presented in this article.


\end{document}